\documentclass[graybox,natbib,nosecnum]{svmult}
\bibpunct{(}{)}{;}{a}{}{,} 

\pdfoutput=1   

\usepackage{mathptmx}       
\usepackage{helvet}         
\usepackage{courier}        
\usepackage{type1cm}        

\usepackage{makeidx}         
\usepackage{graphicx}        
\usepackage{multicol}        
\usepackage[bottom]{footmisc}
\usepackage[normalem]{ulem}	
\usepackage{hyperref}  

\usepackage{multirow}  
\usepackage{wasysym}  

\usepackage{soul}   

\def\Mjup{\hbox{$\mathrm{M}_{\jupiter}$}}
\def\Rjup{\hbox{$\mathrm{R}_{\jupiter}$}}
\def\Mearth{\hbox{$\mathrm{M}_{\oplus}$}}
\def\Rearth{\hbox{$\mathrm{R}_{\oplus}$}}

\newcommand{\hbindex}[1]{\hl{#1}\index{#1}}  

\makeindex             

\begin{document}

\title*{Populations of Extrasolar Giant Planets from Transit and Radial Velocity Surveys}
\titlerunning{Populations of EGP from Transit and RV surveys}
\author{Alexandre Santerne}
\authorrunning{A. Santerne}
\institute{Alexandre Santerne \at Aix Marseille Univ, CNRS, CNES, LAM, Marseille, France, \email{alexandre.santerne@lam.fr}}
%
%
\maketitle

\abstract{Transit and radial velocity surveys have deeply explored the population of extrasolar giant planets, with hundreds of objects detected to date. All these detections allow to understand their physical properties and to constrain their formation, migration, and evolution mechanism. In this chapter, the observed properties of these planets are presented along with the various populations identified in the data. The occurrence rates of giant exoplanets, as observed in different stellar environment by various surveys are also reviewed and compared. Finally, the presence and properties of the giant exoplanets are discussed in the regards of the properties of the host star. Over this chapter, the observational constraints are discussed in the context of the dominant planet formation, migration and evolution scenarios.}

\section{Introduction}
Two decades of exploration of extrasolar \hbindex{giant planets} (hereafter EGPs) with \hbindex{radial velocity} and \hbindex{transit} surveys, both from the ground and from space, have revolutionised our view on giant planets, in comparison with the solar system. At a time when Earth-sized exoplanets are discovered in the habitable zone of their star, many questions regarding the formation, migration, and evolution of EGPs are not yet fully understood. For instance, their dominant formation mechanisms is still debated: either by core-accretion \citep[e.g.][]{2012A&A...547A.112M} or disk instability \citep[see][for a review]{2017PASA...34....2N}. The physical process causing the inflation of giant planets is also unclear \citep{2014prpl.conf..763B}. Even the definition of what is an EGP, with respect to brown dwarfs is actively discussed \citep{2011A&A...532A..79S, 2014prpl.conf..619C, 2015ApJ...810L..25H}. 

Nevertheless, hundreds of EGPs have been discovered and well characterised thanks to photometric surveys like Super-WASP \citep{2006PASP..118.1407P} and HATNet \citep{2004PASP..116..266B} from the ground and \textit{CoRoT} \citep{2006cosp...36.3749B} and \textit{Kepler} \citep{2009IAUS..253..289B} from space, as well as spectroscopic surveys like with the CORALIE and HARPS \citep{2011arXiv1109.2497M}, the SOPHIE \citep[e.g.][]{2016A&A...588A.145H}, and the Lick and Keck \citep[e.g.][]{2005PThPS.158...24M} instruments and observatories (for a more complete list of instruments and surveys, see the corresponding sections of this book). All these discoveries bring important insights into giant planet formation, migration and evolution. This chapter highlights the main interpretation of all the EGP discoveries, starting with a description of the different populations of EGP, then the occurrence rates of EGP as determined in different stellar environments, the relation between the presence and properties of EGPs with respect to their host star, and finally the conclusions.

\section{Different Populations of EGP} 
\label{sect:popEGP}

\begin{figure}[h]
\includegraphics[width=\textwidth]{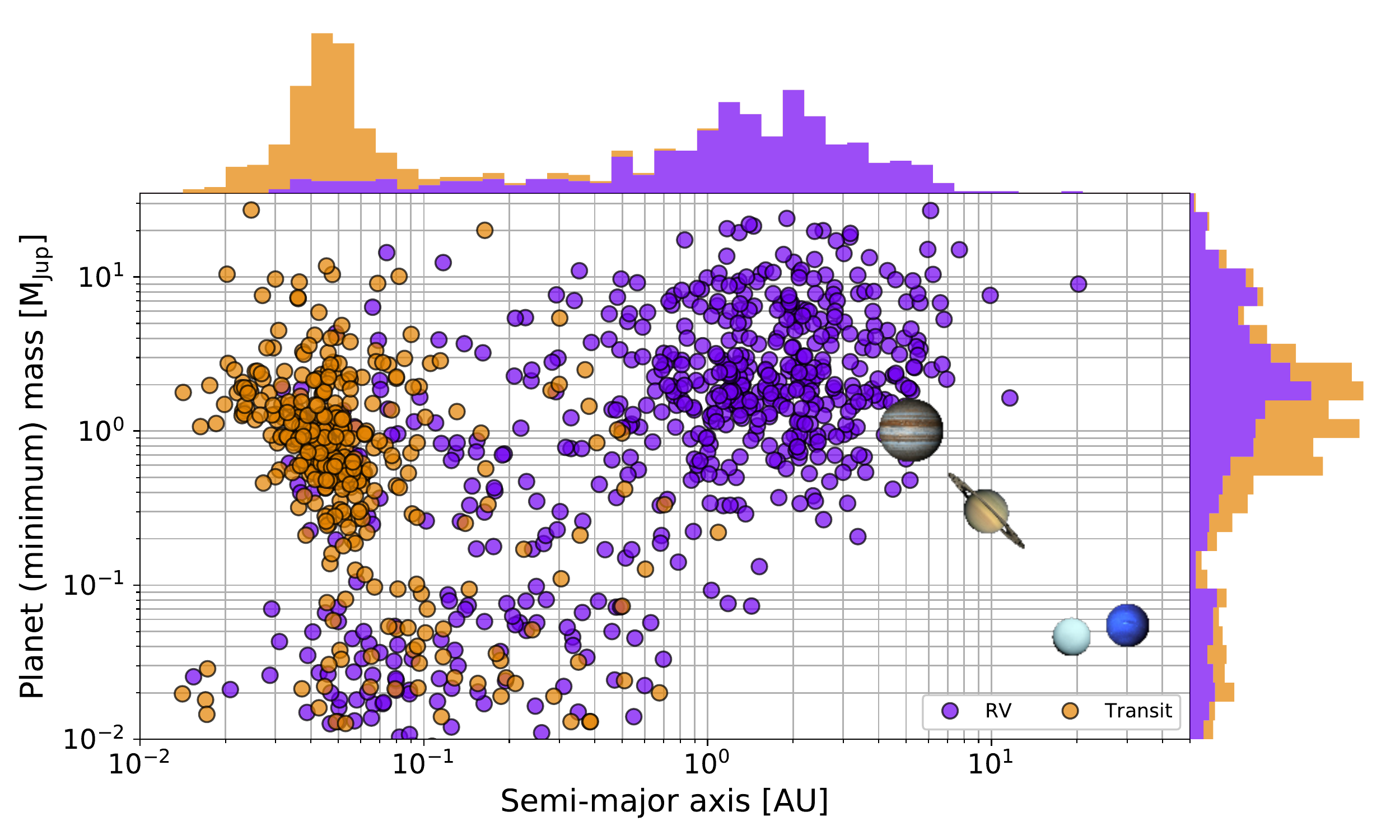}
\caption{Extrasolar planets discovered to date (source: NASA Exoplanet Archive) by the transit (orange marks) and radial velocity (violet marks) surveys. The planetary mass (sky-projected minimum mass in case of RV planets) is shown as function of the semi-major axis of the orbit. The top and right histograms represent the raw distribution of extrasolar planets in their semi-major axis and (minimum) mass, respectively. This reveals the two main populations of EGP and a transitional population where few objects have been found. Solar system planets are also present for comparison.}
\label{santerne-fig1}
\end{figure}

To date, nearly 1000 EGPs have been detected and characterised by photometric and spectroscopic surveys. They are displayed in the Fig. \ref{santerne-fig1} together with their distribution. From all these detections, there is a clear limit between giant planets and lower-mass, Neptune-like planets at about 0.1 -- 0.2 \Mjup\ (about 30 -- 60 \Mearth). In this regime, very few objects have been detected either by transit or RV. This cannot be explained by an observational bias. This lower-limit in mass for the giant planet is supported by the threshold at which the planetesimal starts the \hbindex{runaway accretion} and also open a gap in the disk changing their \hbindex{migration} from type I and type II \citep[e.g.][]{2017Icar..285..145C}. 

The upper-limit in mass, arbitrarily set at $\sim$ 30\Mjup in Fig. \ref{santerne-fig1} corresponds to the one used by the NASA exoplanet archive \citep{2013PASP..125..989A}. 

\subsection{Hot Jupiter}
Planets more massive than 0.1 -- 0.2 \Mjup\ are clearly distributed in two main clusters. The first cluster is very close to the star, with semi-major axis of less than 0.1 AU (equivalent to about 10 days for sun-like stars). These are the so-called \hbindex{hot Jupiters} as these planets are highly irradiated by their host star. The most typical member of this population is 51 Peg b \citep{1995Natur.378..355M}. This population of hot Jupiters has been deeply explored by ground-based photometric surveys with hundreds of detections. They are the easiest planets to detect, even within reach of amateur facilities \citep[e.g.][]{2014EPSC....9..188S}. 
 
\subsection{Temperate/Cold Giants}
The second cluster is located at much larger separation, starting at about 0.4 AU (about 100 days for sun-like stars) up to several AU. This population is composed by temperate to \hbindex{cold giants} and includes planets like Jupiter. Transit surveys from the ground are not very sensitive to planets with periods greater than about 10 days and are not able to probe this population. Only space-based photometry lasting several years like \textit{Kepler} was able to detect giant planets at a few hundreds days, but given their low transit probability (less than 0.5\% at 1AU for a sun-like star), the number of temperate/cold giants detected in transit is small \citep{2016AA...587A..64S}.

Note that the lack of planets detected beyond the orbit of Saturn is only due to observational bias. Two decades of spectroscopic observations allows for the detection of planets with periods up to typically 20 years. As a consequence, planets with orbital period longer than Saturn ($\sim$ 29 years) present only partial orbits in the data which is therefore poorly determined for now. Radial velocity observations in the next decades will likely extend this population of planets towards larger separation, if these planets are common. Planets less massive than Saturn at a few AU are also not yet detected because of instrumental limitations. 

\subsection{Period-valley giants}
Between these two clusters resides a transition where relatively few EGP have been found. This transition is known as the \hbindex{period valley}, first identified by \citet{2003A&A...407..369U} based on radial velocity detections. If this transition population is clear in spectroscopic surveys \citep[see also][]{2007ARA&A..45..397U}, it was first unconfirmed by the transit detections of the \textit{Kepler} mission \citep{2012ApJS..201...15H, 2013ApJ...766...81F}, indicating only one, continuous population of EGPs. This period-valley was however confirmed in the \textit{Kepler} detections by \citet{2016AA...587A..64S} after a systematic removal of false-positive contaminations in the sample using the SOPHIE spectrograph. This valley in the period distribution of EGPs is a strong indication that temperate/cold giants and hot Jupiters have different formation and/or migration mechanisms. 

Nevertheless, this period-valley population appears narrower in the \textit{Kepler} data  \citep[only restricted to within 10 -- 20 days][]{2016AA...587A..64S} compared to spectroscopic data (extended between 10 and about 100 days). One reason for this behaviour is that \textit{Kepler} detected several EGPs with orbital periods between 20 and 100 days that belong to multiple planetary systems, like Kepler-9 \citep{2010Sci...330...51H}, Kepler-51 \citep{2014ApJ...783...53M}, Kepler-89 \citep{2013ApJ...768...14W}, and Kepler-117 \citep{2015A&A...573A.124B}. These EGPs might have stopped their migration in the period valley and did not end as hot Jupiters because of their companion, as it is proposed for the pair Jupiter -- Saturn in the solar system \citep{2007Icar..191..158M}. 

Note that the period valley is poorly explored by ground-based transit surveys as their sensitivity drops drastically above 10 days, i.e. at the upper limit of the hot jupiter population. Therefore, the relative weight between the hot-jupiter, the temperate/cold giants and the period-valley populations seen in Fig. \ref{santerne-fig1} is strongly dominated by observing bias.

\subsection{Lack of Very-Short Period EGP} 

Another clear behaviour observed with all the EGP detections is the lack of low-mass, hot, giant planets (less massive than Jupiter, down to super-Earth) orbiting very close to their star (up to 0.04AU, see Fig. \ref{santerne-fig1}). This desert is fully described in \citet{2016A&A...589A..75M}. Two main reasons have been proposed to explain this desert. The first reasons is that low-mass EGPs in the desert are too much irradiated by their host star and consequently, they loose their atmosphere \citep[as it is observed for some hot Jupiters, e.g.][]{2003Natur.422..143V}. Thus, they migrate down in the mass -- period diagramme. The remnants of photo-evaporated hot Jupiters could be short-period \hbindex{super-Earths} \citep{2014ApJ...793L...3V}. This was however recently ruled out by \citet{2017arXiv170400203W} based on the different metallicity distribution of the host stars. 

The second hypothesis for this dearth of short-period EGP is that the distance at which planets stop their migration depends on the planet mass. This would be the case if the magnetospheric cavity in the inner-edge of protoplanetary disks, which is supposed to stop hot Jupiters migration \citep{2010ApJ...708.1692C}, depends on the mass of the disk and indirectly to the mass of the planet \citep{2016A&A...589A..75M}. This would also be the case if hot Jupiters migrated through a high-eccentricity and tidal circularisation, as the minimum distance between planets and their star depends on their Roche limit, which also depends on their mass \citep{2016ApJ...820L...8M}. However, there is evidence against this high-eccentricity migration model for hot Jupiters, as presented in \citet{2016ApJ...825...62S}. Therefore, this dearth of planets is currently not clearly understood.

\subsection{Multiplicity of EGP}

Spectroscopic surveys of hot Jupiters revealed that about half of them have long-period (at several AU) massive companions, within the planetary or stellar regime \citep[e.g.][]{2014ApJ...785..126K, 2016A&A...586A..93N}. Except in the unique case of the WASP-47 system \citep{2015ApJ...812L..18B} where a hot Jupiter is sandwiched by two low-mass planets, this EGP population does not have nearby low-mass planets. They might have low-mass companions at wide separation but current instrumentation can not detect them. This supports the idea that hot Jupiters could have migrated with two main mechanisms: (1) through interaction with the disk (type I or II migration) or (2) by planet-planet interactions which would make a high eccentricity for the hot-jupiter progenitor caused either by the Lidov -- Kozai effect or planet-planet scattering, and before a tidal circularisation \citep[see][for a review]{2014PNAS..11112616F}.

On the other hand, some temperate and cold giants as well as EGPs in the period valley have inner, low-mass planetary companions \citep[see][]{2014ApJ...790..146F}. This is evidence that these planets had a smooth disk migration that preserved the inner planets, unlike the hot Jupiters. As aforementioned, in some circumstances, the presence of the companion could even be the reason for these EGPs to stay cool and prevent them from migrating inwards and becoming hot Jupiters. The upcoming next-generation instruments, like ESPRESSO \citep{2014AN....335....8P}, will be able to further probe the architecture of EGP systems, and reveal their migration mechanism.

\subsection{Radius of EGP} 

The \hbindex{radius of EGPs} that transit in front of their host star can be measured with relatively high precision. The Fig. 2 show all EGP with measured radius as a function of their incident flux. Because the transit method is more sensitive to planet close to their star, most transiting EGPs known to date are highly irradiated. The EGPs receiving more than $10^{9}$~erg.cm$^{-1}$.s$^{-1}$ exhibit a radius up to 2.2~\Rjup\ (hence, 25~\Rearth). With current models of giant planet atmospheres and internal structure (that are calibrated on Jupiter and Saturn), it is not possible to explain such large radius for a gaseous planet, unless they are extremely young \citep{2015A&A...575A..71A}. Several hypothesis are proposed \citep{2014prpl.conf..763B} but they are still debated. Planets receiving an \hbindex{insolation flux} of less than $10^{8}$ erg.cm$^{-2}$.s$^{-1}$ are yet poorly explored. They have orbital period typically longer than a month and require long-duration space-based photometric surveys, like \textit{CoRoT} and \textit{Kepler} to be detected. Nevertheless, the relatively few objects that were detected to date in this low-incident flux regime do not show sign of inflation. Their radii are in the range 0.5 -- 1.2~\Rjup\ (equivalent to about 6 -- 13~\Rearth). 

\begin{figure}[h]
\includegraphics[width=\textwidth]{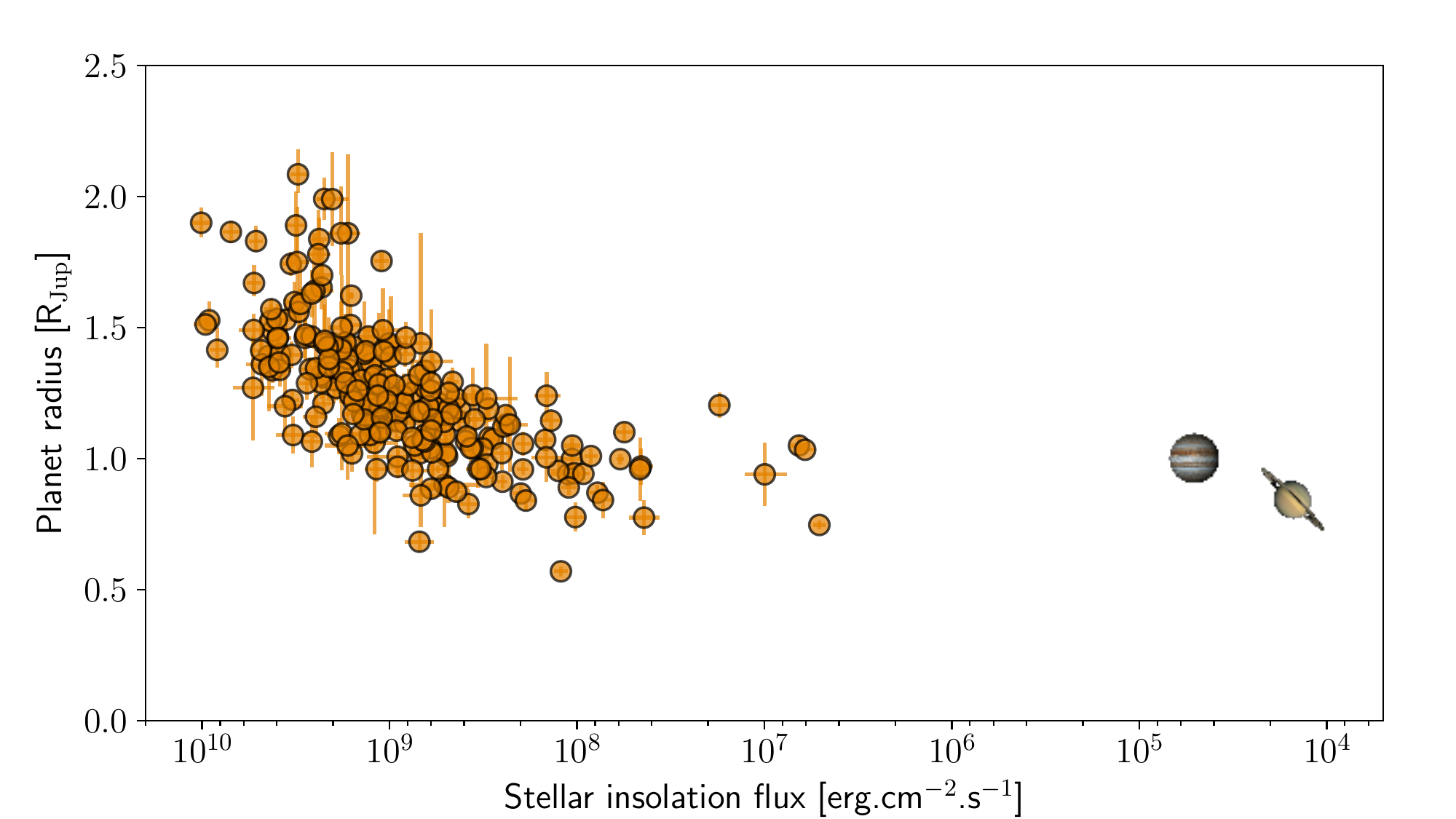}
\caption{Radius of EGP as function of their stellar insolation flux. Jupiter and Saturn are displayed for comparison.}
\label{santerne-fig2}
\end{figure}

\begin{svgraybox}
As a summary, extrasolar giant planets are objects more massive than about 0.1 -- 0.2 \Mjup. The upper-limit in mass is a few tens of \Mjup but is unclear as the distinction between giant planets and brown dwarfs is still debated. In this regime of planetary mass resides two main populations of planets: the hot Jupiters, with orbital separation of less than 0.1 AU (about 10 days of orbital period) and the temperate/cold giants with semi-major axis greater than about 0.4 AU (about 100 days of period). In the transition, the so-called period valley, relatively few planets have been found. The RV and transit detections have also unveil a clear desert of short-period low-mass giants, where no planet has been found. This desert might be the result of a dramatic evolution or a migration of giant exoplanets. Finally, most of the hot Jupiters only have wide-separation more massive companions while warm to cold giants might have inner, low-mass planetary companions, as detected with current instrumentation.
\end{svgraybox}

\section{Occurrence Rates of EGPs in Different Stellar Populations} 
\label{sect:occEGP}

Transit and RV surveys with well defined and characterised stellar samples can be used to derive the occurrence rates of EGPs. One of the main challenges to derive occurrence rate is to correctly estimate the bias inherent of each technique or survey. In the case of the transit surveys, which have targeted relatively faint stars, another challenge is to characterise precisely the observed stellar sample \citep[see][in the case of \textit{Kepler} and \textit{CoRoT}, respectively]{2014ApJS..211....2H, 2016A&A...595A..95D}.

\hbindex{Occurrence rates} of EGPs have been derived based on different surveys that observed different stellar populations across the Galaxy. The values are reported in Table \ref{santerne:tab1} and discussed below. The most up-to-date value of each survey is also displayed in Fig. \ref{santerne-fig3} for comparison. For clarity of the plot, only values with a relative precision better than 50\% are shown.

\subsection{The Solar Neighbourhood}

For now, only RV experiments substantially surveyed the \hbindex{solar neighbourhood} and were used to determine the occurrence rates of EGPs. The two main surveys of exoplanets in RV are the Californian Planet Search (CPS) and the Geneva-lead survey, initiated by \citeauthor{2005PThPS.158...24M} and \citeauthor{2011arXiv1109.2497M}, respectively.

\runinhead{The CPS survey} mostly used the Keck and Lick telescopes with their respective high-resolution iodine-cell based spectrographs. Some observations were also performed with the Anglo-Australian Telescope. The occurrence rate estimates were first published in \citet{2005PThPS.158...24M}, and subsequently in \citet{2008PASP..120..531C}, \citet{2010Sci...330..653H}, and \citet{2012ApJ...753..160W}. The various studies used different sample selections. The latter and most up-to-date study probed the solar neighbourhood with a magnitude-limited sample (V $<$ 8) focused on FGK dwarfs. In this sample, 836 stars were observed with a least 5 epochs which is sufficient to detect EGPs with short orbital periods. They defined hot Jupiters as EGPs with periods of less than 10 days and more massive than mass 0.1\Mjup. Ten hot Jupiters were found in their sample, which corresponds to an occurrence rate of 1.2$\pm$0.38\% (see Table \ref{santerne:tab1}). No occurrence rate was specifically derived for the valley period and temperate/cold giant populations, based on the CPS survey. \citet{2010Sci...330..653H} reported six planets more massive than 30\Mearth\ with orbital periods less than 50 days found in a sub-sample of 122 FGK dwarfs, leading to an occurrence rate in this domain of 4.9$\pm$2.0\%.

\runinhead{The Geneva-lead survey} used both the CORALIE and HARPS spectrographs. The surveys is a volume-limited sample of southern stars up to 50pc focusing on low-activity FGK dwarfs. It is fully described in \citet{2011arXiv1109.2497M} and in \citet{2011A&A...533A.141S}. In total 822 stars were observed by at least one instrument. A total of 155 planets and 6 candidates were detected in this sample, among which 89 have a minimum mass above 50\Mearth\ and are considered as EGP. Correcting for the survey completeness, it gives an overall occurrence rate of EGP within 10 years at the level of 13.9$\pm$1.7\% \citep{2011arXiv1109.2497M}. Selecting only hot Jupiters, valley-period giants, and temperate giants (within 400 days to allow comparison with the \textit{Kepler} transit survey), it gives occurrence rates of 0.83$\pm$0.34\%, 1.64$\pm$0.55\%, and 2.90$\pm$0.72\% as derived in \citet{2016AA...587A..64S}, see also Table \ref{santerne:tab1}. Note that a first attempt to estimate the occurrence rate of EGP in the solar neighbourhood was done with the ELODIE (northern) spectrograph, leading to a low-precision value of 0.7$\pm$0.5\% for hot Jupiters \citep{2005ESASP.560..833N}.

\begin{svgraybox}
In the solar neighbourhood, the occurrence rate of hot Jupiters has been measured by two slightly different surveys at the level of about 1\%.
\end{svgraybox}

\subsection{The Kepler Field}

The \textit{Kepler} space telescope surveyed about 200 000 stars during four years \citep{2009IAUS..253..289B}. The field of view was the same during the prime mission. This field is located about 10 degrees above the galactic plane. Most of the FGK dwarf targets observed in this field are located at several hundreds of parsec away, up to a few kiloparsec. 

Nearly 4700 planet candidates were detected \citep{2016ApJS..224...12C}, among which more than 2200 are confirmed or validated \citep[e.g.][]{2014ApJ...784...45R, 2016ApJ...822...86M}. Therefore, this photometric surveys is a gold mine for statistical analysis, which was actually the prime objective of the mission \citep[e.g.][]{2014PNAS..11112647B}. Statistical properties of exoplanets, including of EGPs, were already derived with the first month (out of 48 in total) of data only \citep{2011ApJ...728..117B}. It was subsequently revised and improved as more data were collected and more candidates were detected. Occurrence rates of exoplanets based on the \textit{Kepler} detections were discussed in numerous papers (see \url{https://exoplanetarchive.ipac.caltech.edu/docs/occurrence_rate_papers.html} for an updated list). In the particular case of EGPs, the occurrence rate was first derived in \citep{2012ApJS..201...15H} for planets up to 0.25AU based on the first four months of data. In this paper, the occurrence rate of hot Jupiters (defined as planets with a radius in the range 8 -- 32\Rearth\ and orbital period of less than 10 days) was estimated to 0.4$\pm$0.1\%. However, because eclipsing low-mass stellar companion can perfectly mimic the transit of an EGP \citep[the so-called \hbindex{false positives}][e.g.]{2012AA...545A..76S}, the reliability of the transit detections to be \textit{bona-fide} exoplanets has to be taken into account. Correcting this bias using spectroscopic ground-based observations with the SOPHIE instrument, \citet{2012AA...545A..76S} re-evaluated the occurrence rate of hot Jupiters (defined as planets with a transit depth in the range 0.4\% -- 3\% and orbital periods less than 10 days) in the \hbindex{\textit{Kepler} field} to be 0.57$\pm$0.07\% also based on the first 4 months of \textit{Kepler} data. Although the two values are fully consistent, the main difference between these two studies is the selection criteria for hot Jupiters and the fact that in the latter analysis, all false positives were screened out with ground-based spectroscopy. 

The false-positive rate of \textit{Kepler} EGPs turned out to be significantly higher \citep[at the level of 35\% for EGP within 25 days detected based on the first 16 months of \textit{Kepler} data; ]{2012AA...545A..76S} than previously estimated \citep[at the level of 5\%][]{2011ApJ...738..170M}. Therefore, to derive more reliable occurrence rates, \citet{2013ApJ...766...81F} simulated the population of false positives, compared it with the population of planetary candidates, and derived occurrence rates for all populations of planets detected in the \textit{Kepler} field, up to about 400 days of orbital periods. For EGP, considered as planets with a radius in the range 6 -- 22\Rearth, they reported occurrence rates of 0.43$\pm$0.05\%, 1.56$\pm$0.11\%, and 3.24$\pm$0.25\% for the hot Jupiter, period-valley, and temperate EGPs, respectively (see Table \ref{santerne:tab1}). 

Extending the sample of \citet{2012AA...545A..76S} to include all EGP up to 400 days of period and detected in the four years of the prime mission (from DR24), \citet{2016AA...587A..64S} observed 125 EGP candidates with the SOPHIE spectrograph to screen out impostors. They finally reported a false-positive rate at the level of 55\% over the whole sample and derived occurrence rates, cleaned from false positives, of 0.47$\pm$0.08\%, 0.90$\pm$0.24\%, and 3.19$\pm$0.73\% for the hot jupiter, period-valley, and temperate giant populations. These values are fully consistents with those derived by \citet{2013ApJ...766...81F}, except for the period-valley EGP where the false-positive contamination was underestimated in the latter study.

\begin{svgraybox}
In the \textit{Kepler} field, the occurrence rate of hot Jupiters has been estimated by different teams using different approaches to correct for the false positives. The most up-to-date value is 0.47$\pm$0.08\%, which is about half the one reported in the solar neighborhood.
\end{svgraybox}

\subsection{The CoRoT eyes}

The french-led \textit{CoRoT} space mission \citep{2006cosp...36.3749B} was the first space mission performing a photometric survey of stars to search for transiting exoplanets. The satellite was pointing different fields of view continuously for periods up to 150 days in two directions (the so-called \hbindex{\textit{CoRoT} eyes}): towards the galactic center and galactic anti-center. The stellar samples observed by \textit{CoRoT} were located at several hundreds of parsec, up to a few kiloparsec near the Galactic plane. As a consequence, the mission was probing two different stellar populations, both being different from the solar neightborhood and the \textit{Kepler} field. 

\textit{CoRoT} was initially designed to explore and characterise the populations of small and long-period planets that are out of reach from ground-based surveys. Deriving the occurrence rates of planets with precision requires to understand the various bias introduced by the different detection teams across Europe \citep{2005A&A...437..355M} as well as the photometric and spectroscopic follow-up process. Moreover, the stellar populations were poorly characterised, hence limiting the precision of occurrence rates. As a consequence, compared to \textit{Kepler}, relatively few occurrence rate studies have been attempted. 

\citet{2012AA...543A.125G} performed a massive spectroscopic survey of \textit{CoRoT} targets in a few pointings toward the galactic anti-center to better characterise the stellar population and derived a first estimate of the occurrence of hot jupiters of 0.4$\pm$0.2\%. This value was later updated by \citet{2013Icar..226.1625M} to 1.0$\pm$0.3\% to include more detections from both galactic directions.

Re-analysing the 5.5 years of the \textit{CoRoT}/Exoplanet programme with an homogeneous transit detection pipeline and selecting the candidates to minimise the bias introduced by the follow-up observations, \citet{2018arXiv180507164D} derived final values for the occurrence rates of hot jupiters in each of the CoRoT eyes. They reported values of 0.95$\pm$0.26\% and 1.12$\pm$0.31\% for the center and anticenter fields, respectively. For EGP with orbital period in the range 10 -- 100 days, the rate is estimated, based on relatively few detections, to 1.53$\pm$1.1\%.

\begin{svgraybox}
In both of the two \textit{CoRoT} eyes, the occurrence rate of hot Jupiters has been estimated to be at the level of about 1\%, in agreement with the value observed in the solar neighborhood.
\end{svgraybox}

\subsection{Ground-based transit surveys}

Many photometric surveys have been performed from the ground with the sensitivity to detect hot Jupiters in many different stellar populations. However, photometric data are highly heterogenous because of variable sky conditions, hence strongly limiting the accurate determination of the survey completeness. The bias introduced by the candidate selection and the follow-up are also challenging to estimate. This explains why the large photometric ground-based surveys, targeting the whole sky, like SuperWASP \citep{2006PASP..118.1407P} and HATNet \citep{2004PASP..116..266B}, have not derived yet occurrence rates. Although this would be extremely interesting, this represents a titanic work. 

Relatively modest \hbindex{ground-based photometric surveys} have been performed on very specific fields of view, allowing to derive estimates of the occurrence rates of the shortest period giant planets. This is the case of the following surveys:
\begin{itemize}
 \item The OGLE survey that observed stars in Carina and the galactic bulge \citep{2006AcA....56....1G}. They derived the occurrence rate of EGPs within 1 -- 3 days and 3 -- 5 days of orbital period. Computing the value for the period range 1 -- 5 days and assuming pure Poisson noise, the occurrence rate is 0.88$\pm$0.39\%. 
 \item The SWEEPS survey \citep{2006Natur.443..534S} targeted the galactic bulge and reported an occurrence rate of EGP within 4.2 days of $0.4^{_{+0.4}}_{^{-0.2}}$\%. 
 \item The Deep MMT survey targeted a field of view in the vicinity of the open cluster M37 \citep{2009ApJ...695..336H}. They reported a 95\% confidence upper-limit on the occurrence rate of EGP within 5 days of orbital period and larger than 1~\Rjup at the level of 3.2\% and 8.3\% for field and cluster member stars.
 \item The SuperLupus survey targeted stars in the Lupus constellation and reported an occurrence rate of hot jupiters at the level of $0.1^{_{+0.27}}_{^{-0.08}}$\% \citep{2011ApJ...743..103B}.
 \end{itemize} 

These four surveys were targeting relatively faint stars, most of them being fainter than Rmag $\sim$ 15 and up to 23. The characterisation of the system with spectroscopic mean requires large telescopes for such faint stars and is thus limited. The number of candidates detected by these surveys is extremely low. This is even lower when considering \textit{bona-fide} EGPs (up to 5 EPGs for the OGLE survey). These occurrence rates are therefore based on small-number statistics and should be interpreted with cautious.

\subsection{In Open Clusters}

\hbindex{Clusters} are pristine targets to test formation and evolution scenarios of exoplanets in different environments than field stars. They also have a well constrained age and metallicity. They have been surveyed for more than a decade with no detection \citep[e.g.][]{2000ApJ...545L..47G, 2009ApJ...695..336H}. Only recently, the first EGPs have been detected in clusters \citep{2012ApJ...756L..33Q}. Based on three detections in radial velocity out of 66 stars in the open cluster M67, \citet{2016AA...592L...1B} reported an occurrence rate of hot Jupiter at the level of $5.6^{_{+5.4}}_{^{-2.6}}$\% and $4.5^{_{+4.5}}_{^{-2.5}}$\% for the single stars and the full sample, respectively. Although these values are much higher than any other estimate, they are also based on very small number statistics, as reflected by their large uncertainties. They should also be interpreted with cautious.

\subsection{The Brown-Dwarf Desert}

\hbindex{Brown dwarfs} are sub-stellar objects with a mass greater than about 10 -- 20 \Mjup, although this is still highly debated \citep{2014prpl.conf..619C}, and up to about 80 \Mjup. These objects are extremely rare in orbit around a solar-type stars within a few AU. This is the so-called brown-dwarf desert \citep[e.g.][]{2000PASP..112..137M}. Based on radial velocity survey, the occurrence rate of brown dwarfs within 3 AU has been estimated to be as low as $0.1^{_{+0.2}}_{^{-0.1}}$\% \citep{2006ApJ...640.1051G}. Using the \textit{Kepler} data and the SOPHIE spectrograph, \citet{2016AA...587A..64S} slightly improved this value to 0.29$\pm$0.17\% for brown dwarfs companion of FGK stars with an orbital period of less than 400 days. This value is consistent with the one found by \citet{2015A&A...584A..13C} of $\sim$0.2\%, for objects within 10 days of orbital period, based on the \textit{CoRoT} detections. Although these estimates are based on small number statistics they reveal that the occurrence rate of companion brown dwarfs in the desert is 5 -- 15 times lower than for EGP in the same stellar sample \citep{2015A&A...584A..13C, 2016AA...587A..64S}.

\subsection{Comparison of the Occurrence Rates}

\begin{figure}[b]
\includegraphics[width=\textwidth]{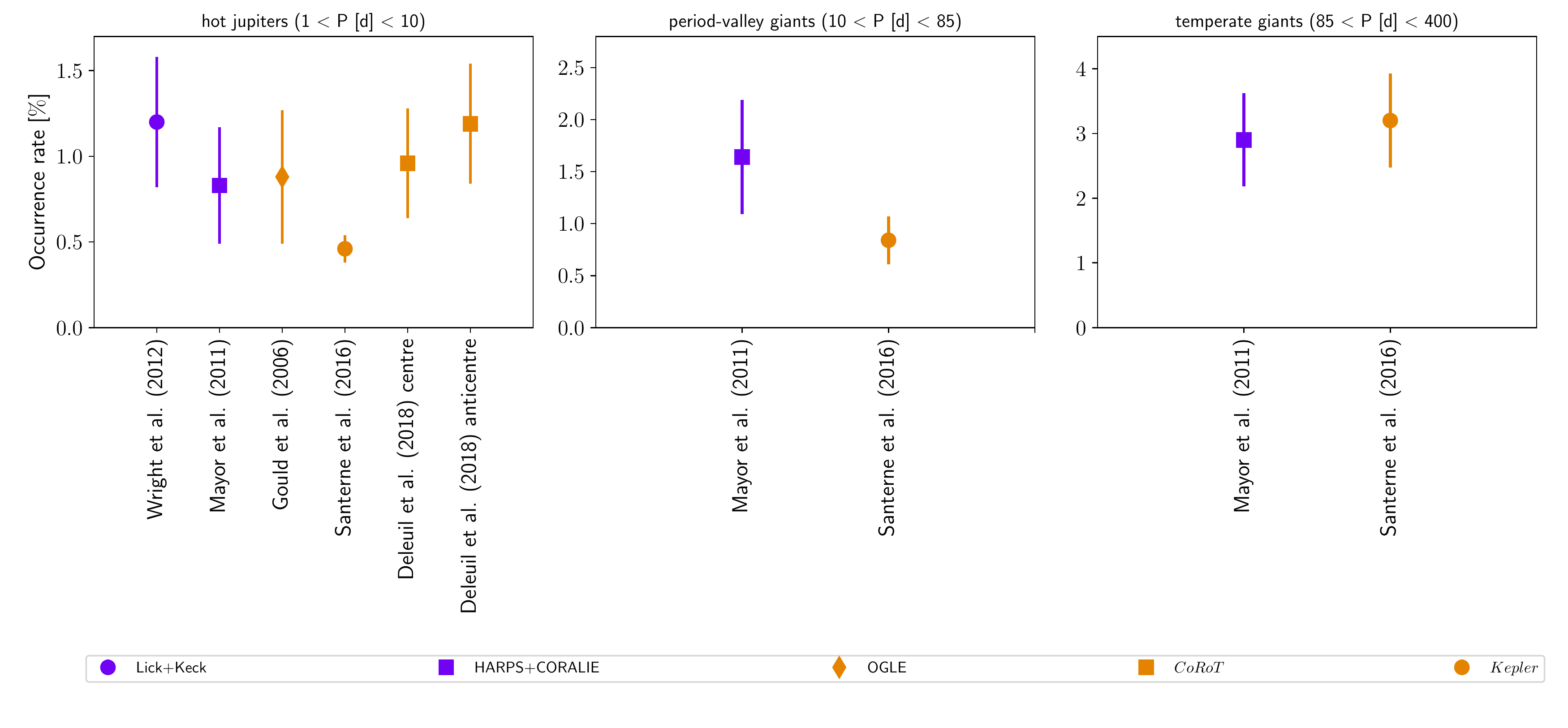}
\caption{Occurrence rates of the different populations of EGP (left: hot Jupiter ; middle: period-valley giants ; right: temperate giants) by different surveys. Violet marks are for RV surveys and orange marks are for photometric surveys. Note the different scale between the three panels. Adapted from \citet{2016AA...587A..64S}.}
\label{santerne-fig3}       
\end{figure}

With two decades of exploration of EGPs, their occurrence rate has been measured by various spectroscopic and photometric surveys in different stellar environments, from the solar neighbourhood, to open clusters. If significant differences are found in these rates between different stellar fields, it will give important insights on the impact of the stellar properties to the formation of EGPs. Comparing the occurrence rates of hot Jupiters in the various surveys aforementioned, it is remarkable that all but the \textit{Kepler} field reported a mean value of 1.00$\pm$0.14\%. In the \textit{Kepler} field, the observed value is nearly half, of 0.47$\pm$0.08\%. Under the assumption that the occurrence rate of hot Jupiter is the same in the solar neighbourhood, the OGLE survey and in both of the \textit{CoRoT} eyes, then the difference with the \textit{Kepler} field is at more than 3-$\sigma$.

While the \textit{Kepler} field was initially thought to be of lower \hbindex{metallicity} (see next section for the impact of the stellar metallicity of the EGP formation) compared to the other surveys \citep[and references therein]{2016AA...587A..64S}, massive spectroscopic survey of \textit{Kepler} stars reported it to be consistent with a solar metallicity and furthermore, compatible with the other surveys \citep{2014ApJ...789L...3D, 2017ApJ...838...25G}. This difference in the rates is therefore unlikely explained by the difference of metallicity of the stellar field. Several other solutions are proposed such as a different \hbindex{stellar multiplicity} rate or some significant bias in the stellar population observed by \textit{Kepler}, but it remains unclear today.

It is interesting to note that while the occurrence rate of hot Jupiter is significantly lower in the \textit{Kepler} field compared to other surveys, this does not seem to be the case for temperate EGPs (see Fig. \ref{santerne-fig3}). The occurrence rates of EGPs in the period valley and of temperate planets, assumed to be within orbital periods of 10d -- 85d and 85d -- 400d (respectively) have been estimated in both the solar neighbourhood based on the HARPS+CORALIE sample and in the \textit{Kepler} field \citep{2011arXiv1109.2497M, 2016AA...587A..64S}. Although the uncertainties are large, the occurrence rates of giant planets at long orbital period are very consistent between the two surveys. For the transition population, if the depletion of EGPs is real, it is not significant in the data. Further investigation are mandatory to fully understand this picture.

\begin{table}[t]
\caption{Occurrence rates of EGP for different ranges of orbital periods from different studies. All values are in percent.}
\begin{center}
\setlength{\tabcolsep}{1mm}
\begin{tabular}{lcccl}
\hline
\multirow{3}{*}{Reference} & Hot Jupiters & Period-Valley Giants & Temperate Giants & \multirow{3}{*}{Survey / Field}\\
 & $P<10$ d & $10<P<85$ d & $85<P<400$ d & \\
 & [\%] & [\%] & [\%] & \\
\svhline
\citet{2005PThPS.158...24M} & 1.2$\pm$0.2 & -- & -- & Keck+Lick+AAT \\
\citet{2005ESASP.560..833N} & 0.7$\pm$0.5$^{(a)}$ & -- & -- & ELODIE \\
\citet{2008PASP..120..531C} & 1.5$\pm$0.6 & -- & -- & Keck \\
\citet{2011arXiv1109.2497M} & 0.83$\pm$0.34$^{(b)}$ &  1.64$\pm$0.55$^{(b)}$  &  2.90$\pm$0.72$^{(b)}$ & HARPS+CORALIE\\
\citet{2012ApJ...753..160W} & 1.20$\pm$0.38 & -- & -- & Keck+Lick \\
\citet{2016AA...592L...1B} & 4.5$^{_{+4.5}}_{^{-2.5}}$  & -- & -- & M67\\
\hline
\citet{2006AcA....56....1G} & 0.88$\pm$0.39$^{(c)}$ & -- & -- & OGLE \\
\citet{2006Natur.443..534S} & 0.4$^{_{+0.4}}_{^{-0.2}}\ ^{(d)}$ & -- & -- & SWEEPS \\
\citet{2009ApJ...695..336H} & $<8.3^{(e)}$ & -- & -- & Deep MMT (M37) \\
\citet{2009ApJ...695..336H} & $<3.2^{(e)}$ & -- & -- & Deep MMT (field stars) \\
\citet{2011ApJ...743..103B} & 0.10$^{_{+0.27}}_{^{-0.08}}$ & -- & -- & SuperLupus\\
\citet{2012ApJS..201...15H} & 0.4$\pm$0.1 & -- & --  & \textit{Kepler}\\
\citet{2012AA...545A..76S} & 0.57$\pm$0.07 & -- & --  & \textit{Kepler}\\
\citet{2013ApJ...766...81F} & 0.43$\pm$0.05  & 1.56$\pm$0.11  &  3.24$\pm$0.25  & \textit{Kepler}\\
\citet{2013Icar..226.1625M} & 1.0$\pm$0.3$^{(f)}$ & -- & -- & All \textit{CoRoT} fields\\
\citet{2016AA...587A..64S} & 0.47$\pm$0.08  &  0.90$\pm$0.24  &  3.19$\pm$0.73 & \textit{Kepler}\\
\citet{2018arXiv180507164D} & 0.95$\pm$0.26 & \multirow{2}{*}{1.53$\pm$1.1} & -- & \textit{CoRoT} center fields\\
\citet{2018arXiv180507164D} & 1.12$\pm$0.31 &  & -- & \textit{CoRoT} anti-center fields\\
\hline
\end{tabular}
\end{center}
The horizontal line separates the values determined by RV surveys (above) from the ones determined by photometric transit surveys (below).\\
$^{(a)}$ Only for periods less than 5 days, but no planet is detected in the sample stars with orbital periods between 5 and 10 days. Therefore, a different period bin, up to 10 days would have provided the same value.\\
$^{(b)}$ Values derived in \citet{2016AA...587A..64S} based on the detections and bias correction provided in \citet{2011arXiv1109.2497M}. The occurrence rate of hot Jupiters slightly differ from the original paper to account only for planets within 10 days of period. The only difference is the planet HD108147b (P=10.89d) that is included here in the population of the period-valley giants and not in the hot Jupiter population.\\
$^{(c)}$ Based on detections up to 5 days. Uncertainty estimated from pure Poisson noise.\\
$^{(d)}$ Based on detections up to 4.2 days.\\
$^{(e)}$ 95\% confidence upper-limit for 1\Rjup\ planets with orbital periods up to 5 days.\\
$^{(f)}$ Update of the value provided in \citet{2012AA...543A.125G} based on the first few \textit{CoRoT} fields.
\label{santerne:tab1}
\end{table}%

\section{Relation with Host Star Properties}
\label{sect:EGPHost}

The environmental conditions in which EGPs form should influence their physical properties that are observed now. To test what are the key ingredients for planet formation, one can use the properties of the host stars as proxy of the initial conditions. The relation between the host star and the EGP properties provides extremely important constraints to understand their formation. The main stellar properties that have been reported so far to significantly influence the formation of EGPs are the metallicity and the mass.

\subsection{The Stellar Metallicity}

A connection between the \hbindex{stellar metallicity} and the presence of an EGP was first identified by \citet{1997MNRAS.285..403G} based on the first four EGPs discovered. They were found to orbit preferentially metal-rich stars. The correlation between the presence of an EGP and the metallicity of the host star was later revised and further characterised as the number of EGPs increased, in particular in \citet{2001A&A...373.1019S, 2003A&A...398..363S} and in \citet{2005ApJ...622.1102F}. This correlation was seen as an observational evidence for the core-accretion mechanism as the primary formation scenario for EGP. Indeed, protoplanetary disks accrete materials to form planetesimal more efficiently if they are rich in metals \citep{1996Icar..124...62P} than if they are depleted in metals. On the other hand, the disk instability scenario was predicting a flat distribution of EGP-host metallicities \citep{2002ApJ...567L.149B}. However, the latest versions of the disk instability mechanism are capable of reproducing this correlation \citep{2015MNRAS.452.1654N, 2017PASA...34....2N}. 

\begin{figure}[h]
\includegraphics[width=\textwidth]{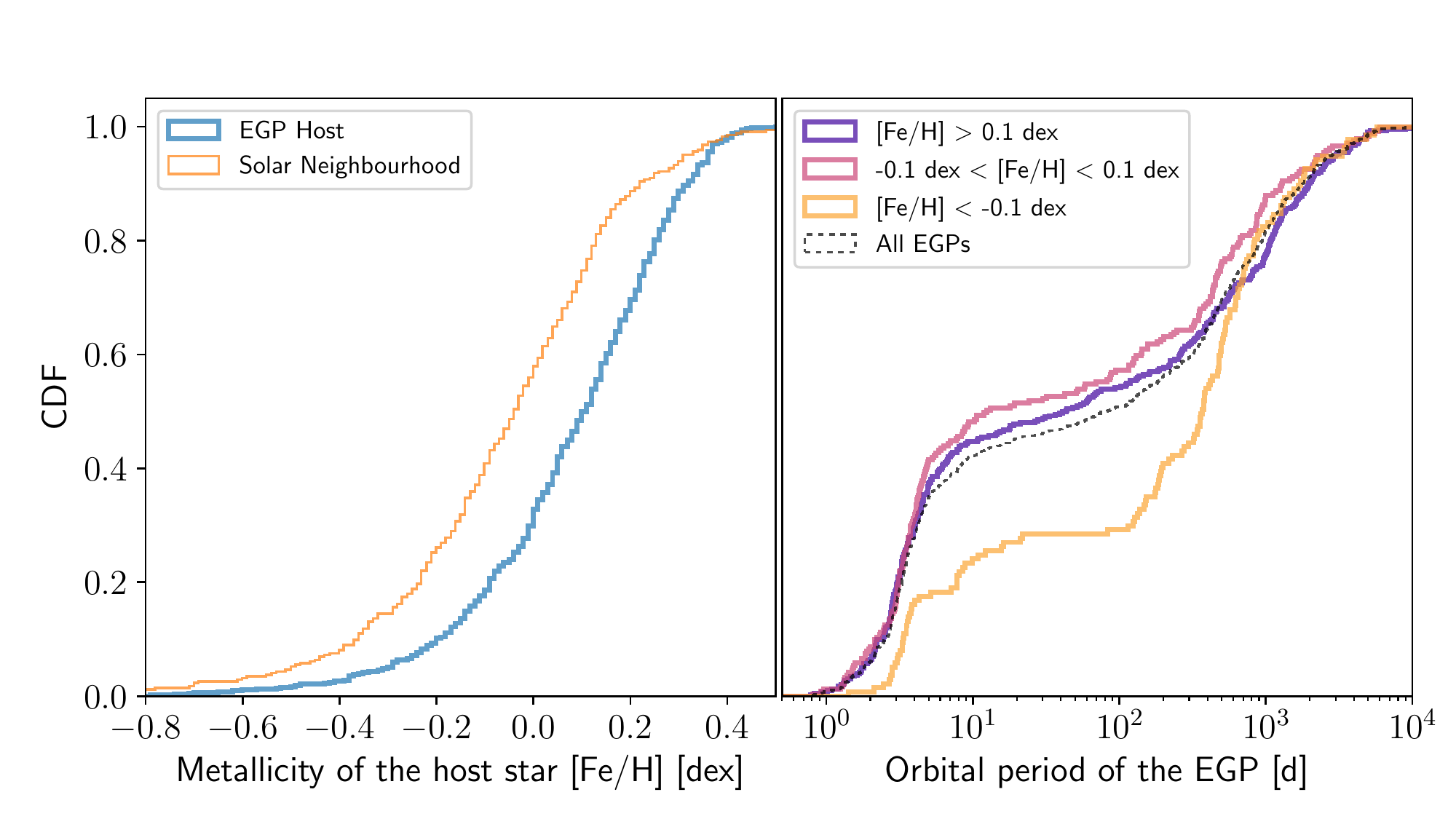}
\caption{(Left) Cumulative distribution functions (CDF) of the metallicity of EGP host stars (thick blue line) and of the solar neighbourhood \citep[thin orange line ;][]{2001MNRAS.325.1365H}. (Right) CDF of the orbital period of EGPs for three populations of host stars: the metal-poor (orange), the solar metallicity (pink), and metal-rich (violet) stars. The dotted line is the CDF for all EGPs. Only those detected by the transit and RV techniques are displayed here. The metallicity data of the host stars are from the Sweet-Cat catalogue \citep{2013A&A...556A.150S}.}
\label{santerne-fig4}       
\end{figure}

Figure \ref{santerne-fig4} displays the cumulative distribution function of the metallicity of all EGP hosts known to date \citep[data from the SweetCat catalog ;][]{2013A&A...556A.150S}, compared to the distribution of metallicity in the solar neighbourhood \citep{2001MNRAS.325.1365H}. While about 50\% of stars in the Sun's vicinity are metal-rich (relative to the Sun value), nearly 80\% of EGP hosts are metal rich. This clearly illustrates that EGP prefers to form around metal-rich stars.

When the number of EGPs was high enough, this correlation was characterised to be reproduced by a power law of the form:

\begin{equation}
\mathcal{P}({\rm planet}) = \alpha [Fe/H]^\beta\ ,
\end{equation}
with $\alpha = 0.03$ and $\beta = 2.0$ \citep{2005ApJ...622.1102F}. This functional form was then revisited in \citet{2010PASP..122..905J} and \citet{2013A&A...551A.112M} to account for the role of the stellar mass (see next section) and discuss the shape of the distribution at low-metallicity values. The number of EGPs known today is still too low to precisely constrain the form of the correlation in the low metallicity regime.

The metallicity of the host star does not only impact the presence of planets, it also shape their period distribution. As pointed out by \citet{2013A&A...560A..51A}, EGP in metal-poor systems are found at much larger separation than those in the metal-rich conterpart. This is illustrated in Fig. \ref{santerne-fig4} where EGPs orbiting stars with [Fe/H] $<$ -0.1 dex clearly have much longer orbital periods than those orbiting stars more rich in metals. This is explained by the fact EGPs still form around metal-poor stars, but slower than in metal-rich system. As a consequence, their migration is also less efficient and they are observed close to their formation place.

\subsection{The Stellar Mass}

The mass of the central star is known to impact the formation of EGPs. Indeed, the mass of the protoplanetary disk scale with the mass of the star. Hence, the more massive disks have the potential to form more efficiently massive planets, compared to low-mass disks. On one end, extremely few giant planets have been found so far to orbit the low-mass \hbindex{M dwarfs} \citep[among the few cases, there are GL 876 b and Kepler-42 b ;][respectively]{1998A&A...338L..67D,2012AJ....143..111J}. Therefore, protoplanetary disks surrounding M dwarfs are not massive enough to form efficiently EGPs.

On the other end, massive stars form with massive disks and should form efficiently EGPs. This is however challenging to explore as O, B, A, and early F stars have few lines in the optical and rotate fast hence strongly limiting the radial velocity technique. The approach employed by e.g. \citet{2007ApJ...665..785J} was to search for EGPs orbiting \hbindex{retired A stars}, i.e. giant stars for which precise RVs can be obtained. While a relatively large number of EGPs were found around evolved stars, determining the mass of the host star accurately is challenging without the use of asteroseismology. Indeed stellar evolutionary tracks from a wide range of stellar mass converge towards the so-called giant branch. The mass determination of the retired A stars, then used to quantify the impact of the stellar mass on the formation of EGP \citep{2010PASP..122..905J}, is controversial \citep{2011ApJ...739L..49L, 2013ApJ...774L...2L}. Nevertheless, EGPs are found more abundant around K giants, than around solar-like stars \citep{2010ApJ...709..396B}, hence confirming the trends observed with M dwarfs. It is however expected that a cut-off high mass exists, around which EGP do not form, e.g. because the star evolves too fast off of the main-sequence, or because of intense stellar winds.

\begin{svgraybox}
The properties of the central star have a direct impact of the formation of EGP. The most important is the host metallicity. EGPs form preferentially around metal-rich stars. Those forming around metal-poor stars have a wider separation than their metal-rich counterparts. The mass of the central star also play an important role on the formation of EGP. The more massive the host star is, the more likely they form EGP. As a consequence, EGPs are very rare around M dwarfs.
\end{svgraybox}

\section{Conclusions}
\label{sect:EGPConc}

Extrasolar giant planets have been explored for more than two decades with transit and radial velocity surveys, leading to several hundreds of detections. From this large sample of objects, it is possible to infer about their physical properties and to identify different populations. Both are providing important constraints for planet formation and migration theories. The EGPs have been found to have a mass greater than about 0.1 -- 0.2 \Mjup, which corresponds to the critical mass to start runaway accretion and type II migration. The upper-mass limit of EGPs is unclear and still debated. It would be located between 13 \Mjup, the official limit as defined by the IAU, up to 25 \Mjup. The EGPs that received a moderate stellar irradiation exhibit a radius in the range 0.5 -- 1.2 \Rjup while those extremely close to their star might have radius up to about 2.2 \Rjup.

Among these EGPs detected by transit and RV surveys, three populations can be identified: 
\begin{enumerate}
\item Very close to the host stars (with an orbital period of less than about 10 days) exists a population of EGPs called the hot Jupiters. These planets have an occurrence rate at the level of about 1 \% as observed in the solar neighborhood with the RV surveys (HARPS+CORALIE and Keck+Lick among the main surveys), in both of the \textit{CoRoT} eyes as well as in the OGLE photometric survey. However, this value has been reported to be nearly half in the \textit{Kepler} field for a reason that is still unclear. The formation and migration mechanisms for these EGPs is still debated. Among the most probable scenarios, there is the high-eccentricity with tidal circulation scenario caused by the Lidov -- Kozai mechanism or planet-planet scattering. Another scenario would be a disk-driven migration, although this hypothesis is not clearly compatible with the diverse orbital obliquities of these planets.
\item Much farther out from the host star (with an orbital period greater than about 100 days) exists a population of cool (or temperate) EGPs. These planets have been mainly explored by the RV surveys as their transit probability is extremely low (less than about 1\%) and require long-duration, high-cadence space-based observations like with the \textit{CoRoT} or \textit{Kepler} missions. Their occurrence rate can reach up to about 14\% for EGPs within 10 years of orbital periods. In the period range 85 -- 400 d, covered by both the RV and \textit{Kepler} surveys, their occurrence rate has been estimated at the level of 3 \%. This value is fully consistent between the two surveys, albeit the uncertainties are relatively large. These EGPs likely formed in-situ or had a relatively soft migration from their birth place.
\item In between the hot jupiter and cool/temperate giants resides a transition population of planets. They are called the period-valley giants or even warm giants. They have an orbital period roughly in the range 10 -- 100 days. Their occurrence rate has been reported up to 1.6 \% in the solar neighbourhood and down to 0.9 \% in the \textit{Kepler} field, albeit the uncertainties are large enough to prevent this difference to be statistically significant. The origin of these EGPs is still unclear and they are probably in the tail of the distribution of the other two main populations.
\end{enumerate}

The properties of the host star, proxy of the properties of the protoplanetary disk, has been found to have an impact on the formation of EGPs. The more metal-rich stars are, the more likely they form EGPs. The metal-poor stars still form EGPs but these planets are observed with a much wider orbital separation than those orbiting metal-rich stars. This is explained by the fact the core-accretion mechanism is less efficient to form giant planets in metal-poor disks. Therefore, when EGPs form in metal-poor disk, they need more time to accrete materials and hence migrate less efficiently. The mass of the star also impacts the formation of EGPs: the more massive the central star is, the more likely they formed with EGPs. As a consequence, EGPs are extremely rare in orbit around the low-mass M dwarfs.

More than two decades after the discovery of the first EGP, and at a time when the community interest moves towards small and cool planets, there are still many open questions to explain all the observed properties of the EGPs. For instance, the formation and migration scenario of hot jupiters, the transition between EGPs and brown dwarfs, the reason for the inflated radius of highly-irradiated EGPs, the difference between the occurrence rate of hot Jupiters as observed in the \textit{Kepler} field and the one of all the other surveys, all these questions are still debated. To solve them, more theoretical models have to be developed together with stronger observational constraints.

\section{Cross-References}
\begin{itemize}

\item{Radial Velocities as an Exoplanet Discovery Method}
\item{Transit Photometry as an Exoplanet Discovery Method}
\item{High-Precision Spectrographs for Exoplanet Research: Elodie, Coralie, Sophie and HARPS}
\item{Space Missions for Exoplanet Science with CoRoT}
\item{Space Missions for Exoplanet Science: Kepler / K2}
\item{The SuperWASP and NGTS Exoplanet Surveys}
\item{The HATNet and HATPI Exoplanet Surveys}
\item{Mass-Radius Relations of Giant Planets: The Radius Anomaly and Interior Models}
\item{Exoplanet Occurrence Rates from Transit and RV Surveys}
\item{Formation of Giant Planets}
\item{Extrasolar Planets Population Synthesis}
\end{itemize}

\bibliographystyle{spbasicHBexo}  
\bibliography{Santerne} 

\end{document}